\documentclass[aps,prl,twocolumn]{revtex4-1}

\usepackage{amsmath,amssymb}
\usepackage{color}
\usepackage{graphicx}
\usepackage{mathbbol}

\begin{document}

\title{Chaotic Blowup in the 3D Incompressible Euler Equations on a Logarithmic Lattice}

\author{Ciro S. Campolina}\email{sobrinho@impa.br}
\author{Alexei A. Mailybaev}\email{alexei@impa.br}

\affiliation{Instituto Nacional de Matem\'atica Pura e Aplicada---IMPA, 22460-320 Rio de Janeiro, Brazil}

\begin{abstract}
The dispute on whether the three-dimensional (3D) incompressible Euler equations develop an infinitely large vorticity in a finite time (blowup) keeps increasing due to ambiguous results from state-of-the-art direct numerical simulations (DNS), while the available simplified models fail to explain the intrinsic complexity and variety of observed structures. Here, we propose a new model formally identical to the Euler equations, by imitating the calculus on a 3D logarithmic lattice. This model clarifies the present controversy at the scales of existing DNS and provides the unambiguous evidence of the following transition to the blowup, explained as a chaotic attractor in a renormalized system. The chaotic attractor spans over the anomalously large six-decade interval of spatial scales. For the original Euler system, our results suggest that the existing DNS strategies at the resolution accessible now (and presumably rather long into the future) are unsuitable, by far, for the blowup analysis, and establish new fundamental requirements for the approach to this long-standing problem.
\end{abstract}

\maketitle

The existence of blowup (a singularity developing in a finite time from smooth initial data) in incompressible ideal flow is a long-standing open problem in physics and mathematics. Such blowup is anticipated by Kolmogorov's theory of developed turbulence~\cite{frisch1999turbulence}, predicting that the vorticity field diverges at small scales as $\delta\omega \sim \ell^{-2/3}$, while the time of the energy transfer between the integral and viscous scales remains finite in the inviscid limit.
In this context, the blowup would reveal an efficient mechanism of energy transfer to small scales. Similar open problems on finite-time singularities, which are fundamental for the understanding of physical behavior, exist across many other fields such as natural convection~\cite{majda2002vorticity}, geostrophic motion~\cite{pedlosky2013geophysical,constantin1994formation}, magnetohydrodynamics~\cite{biskamp1997nonlinear}, plasma physics~\cite{glassey1986singularity,andreasson2011einstein} and, of course, general relativity~\cite{choquet2009general}.

Besides purely mathematical studies, e.g.,~\cite{beale1984remarks,chae2008incompressible,tao2016finite}, a crucial role in the blowup analysis is given to direct numerical simulations (DNS). The chase after numerical evidence of blowup in the 3D incompressible Euler equations has a long history~\cite{gibbon2008three}. Most early numerical studies were in favor of blowup, e.g., \cite{pumir1992finite,kerr1993evidence,grauer1998adaptive}. But the increase of resolution owing to more powerful computers showed that the growth of small-scale structures may be depleted at smaller scales, even though it was demonstrating initially the blowup tendency~\cite{hou2007computing,grafke2008numerical,hou2009blow}. It is fair to say that, now, there is a lack of consensus even on the more probable answer (existence or not) to the blowup problem. Blowup remains an active area of numerical research~\cite{kerr2013bounds,brenner2016potential,larios2018computational}, but computational limitations are still the major obstacle. See also~\cite{luo2013potentially,elgindi2018finite} for the blowup at a physical boundary, which is a related but different problem. 

Numerical limitations of the DNS can be overcome using simplified models~\cite{uhlig1997singularities,dombre1998intermittency,mailybaev2012renormalization}, which were developed in lower spatial dimensions~\cite{constantin1985simple,okamoto2008generalization} or by exploring the cascade ideas in so-called shell models~\cite{gledzer1973system,ohkitani1989temporal,l1998improved}. The reduced wave vector set approximation (REWA) model introduced in~\cite{eggers1991does,grossmann1996developed} restricted the Euler or 
Navier-Stokes dynamics to a self-similar set of wave vectors. Despite being rather successful in the study of turbulence~\cite{grossmann1992intermittency,biferale2003shell,bohr2005dynamical}, these models 
fall short of reproducing basic features of full DNS for the blowup phenomenon. 

Here, we resolve this problem with a new model that demonstrates qualitative agreement with the existing DNS and permits a highly reliable blowup analysis. The model is formulated in a form identical to the original Euler equations, but with the algebraic structure defined on the 3D logarithmic lattice. We show that the blowup in this model is associated with a chaotic attractor of a renormalized system, in accordance with some earlier theoretical conjectures~\cite{pomeau2005unfinished,greene2000stability,mailybaev2012c,de2017chaotic}; one can also make an interesting connection with the chaotic Belinskii-Khalatnikov-Lifshitz singularity in general relativity~\cite{belinskii1970oscillatory,choquet2009general}. A distinctive property of the attractor is its anomalous multiscale structure, which explains the diversity of the existing DNS results, discloses fundamental limitations of current strategies, and provides new guidelines for the original blowup problem.

\begin{figure*}[t]
\centering
\includegraphics[width=\textwidth]{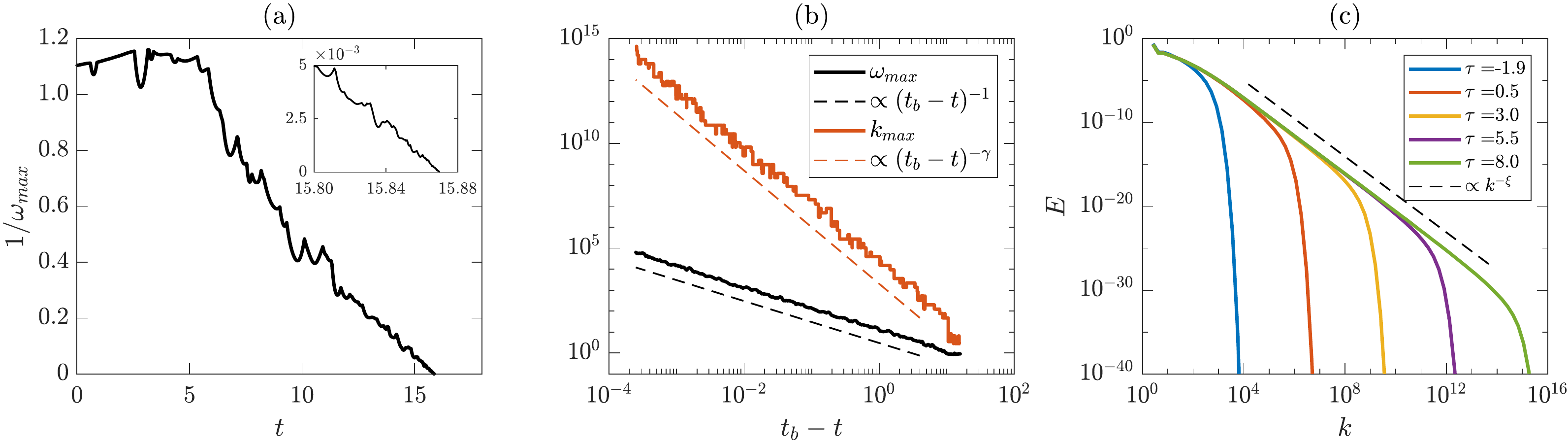}
\caption{(a) Inverse maximum vorticity, $1/\omega_{\max}$, as a function of time; the inset displays an amplified segment very close to the blowup time $t_b = 15.870$. The graph shows deterministic chaotic oscillations; it is not smooth, because the vorticity maximum jumps between nodes of the 3D lattice with increasing time. (b) Evolution of maximum vorticity in log-scale, demonstrating chaotic oscillations around the power law $\propto (t_b-t)^{-1}$, and wave number $k_{\max}$ corresponding to the vorticity maximum, following in average the power law $(t_b-t)^{-\gamma}$ with $\gamma = 2.70$. (c) The energy spectrum, $E(k) = \frac{1}{2\Delta}\sum_{k \le |\mathbf{p}| < \lambda k}|\mathbf{u}(\mathbf{p})|^2$ with $\Delta = \lambda k-k$, in log-scale at different renormalized times $\tau = -\log(t_b-t)$. As $t \to t_b$ corresponding to $\tau \to \infty$, the spectrum develops the power law $E \propto k^{-\xi}$ with $\xi = 3-2/\gamma \approx 2.26$.} 
\label{fig1}
\end{figure*}

\textsc{Model.} Consider the set $\mathbb{\Lambda} = \{ \pm \lambda^{n},\,n\in\mathbb{Z}\}$ of positive and negative integer powers of a fixed real number $\lambda>1$. Then wave vectors $\mathbf{k} = (k_1,k_2,k_3) \in \mathbb{\Lambda}^3$ define a logarithmic lattice in 3D Fourier space. We retain three independent spatial directions, unlike shell or REWA models~\cite{biferale2003shell,eggers1991does} featuring a fixed number of wave vectors per spherical shell. In analogy to the convolution operation, we define
\begin{equation}
(u \ast v)(\mathbf{k}) = \sum_{\substack{\mathbf{p},\mathbf{q} \in \mathbb{\Lambda}^3\\[2pt] \mathbf{p} + \mathbf{q} = \mathbf{k}}} u(\mathbf{p})v(\mathbf{q})  
\label{convolution}
\end{equation}
for complex-valued functions $u(\mathbf{k})$ and $v(\mathbf{k})$. Since the sum is restricted to exact triads on the lattice $\mathbf{k} \in \mathbb{\Lambda}^3$, operation (\ref{convolution}) is nontrivial only for specific values of $\lambda$. We will consider the golden mean, $\lambda = (1+\sqrt{5})/2$, which also appeared in a similar context for shell models~\cite{l1999hamiltonian,gurcan2017nested}. In this case the sum in \eqref{convolution} contains 216 distinct terms originating from the equality $\lambda^{n-1}+\lambda^n = \lambda^{n+1}$ and coupling the wave numbers that differ by $\lambda$ or $\lambda^2$ in each spatial direction. Note that Eq.~(\ref{convolution}) can be seen as a projection of the convolution to the nodes of the 3D logarithmic lattice, which keeps the middle-range interactions. One may expect that the long-range interactions are less important for the blowup problem than, e.g., for the developed turbulence, because very small scales are weakly perturbed in smooth initial conditions. 

Just like the classical convolution, operation \eqref{convolution} is bilinear, commutative and satisfies the Leibniz rule 
\begin{equation}
\partial_j(u \ast v) = \partial_j u \ast v + u \ast \partial_j v \quad \text{for} \quad j=1,2,3,
\label{leibniz}
\end{equation}
where derivatives are given by the Fourier factors, $\partial_j u(\mathbf{k}) = ik_j u(\mathbf{k})$; here $i$ is the imaginary unit. However, operation \eqref{convolution} is not associative, $(u \ast v) \ast w \ne u \ast (v \ast w)$. Nonetheless, it possesses the weaker property 
\begin{equation}
\langle u \ast v, w \rangle = \langle u , v \ast w\rangle,
\label{associativity}
\end{equation}
where $\langle u, v \rangle = \sum_{\mathbf{k} \in \mathbb{\Lambda}^3 } u(\mathbf{k})v^*(\mathbf{k})$ is the scalar product.

In our simplified model, we represent the velocity field as a function $\mathbf{u}(\mathbf{k},t) = (u_1,u_2,u_3) \in \mathbb{C}^3$ of the wave vector $\mathbf{k} \in \mathbb{\Lambda}^3$ and time $t \in \mathbb{R}$. Thus, at each lattice point, $\mathbf{u}$ stands for the corresponding  velocity in Fourier space. Similarly, we define the scalar function $p(\mathbf{k})$ representing the pressure. All functions are supposed to satisfy the reality condition: $u_i(-\mathbf{k}) = u_i^*(\mathbf{k})$. For the governing equations, we use the exact form of 3D incompressible Euler equations
\begin{equation}
\partial_t u_i + u_j \ast \partial_j u_i = -\partial_i p, \quad \partial_j u_j=0,
\label{euler}
\end{equation}
which are now considered on the logarithmic lattice; here and below repeated indices imply the summation. 

The proposed model retains most of the properties of the continuous Euler equations, which rely only upon the structure of the equations and elementary operations such as (\ref{leibniz}) and (\ref{associativity}). These include the basic symmetries: scaling (in a discrete form $\mathbf{k} \mapsto \lambda\mathbf{k}$), isotropy (reduced to the discrete group $\mathsf{O_h}$ of cube isometries \cite{landau2013quantum}, Sec. 93), and spatial translations (given in Fourier representation by $\mathbf{u} \mapsto e^{-i\mathbf{a}\cdot\mathbf{k}}\mathbf{u}$). The system conserves energy $E = \frac{1}{2} \langle u_j, u_j \rangle$ and helicity $H = \langle u_j ,\omega_j \rangle$, where $\pmb{\omega} = \nabla \times \mathbf{u}$ is the vorticity. It also has an infinite number of invariants, which can be interpreted as Kelvin's circulation theorem; see the Supplemental Material (SM)~\cite{SupMat}. Proofs of all these properties are identical to the continuous case.

\textsc{Simulations.}
For numerical simulations, we used the Euler equations in vorticity formulation
\begin{equation}
\partial_t \omega_i + u_j \ast \partial_j \omega_i - \omega_j \ast \partial_j u_i = 0,
\label{vorticity}
\end{equation}
where $\mathbf{u} = \mathrm{rot}^{-1}\pmb{\omega} = i\mathbf{k}\times \pmb{\omega}/|\mathbf{k}|^2$.
Aiming for the blowup study, we consider initial conditions limited to large scales, $\lambda \le |k_{1,2,3}| \le \lambda^3$; see SM~\cite{SupMat} for an explicit form of the initial conditions. Equations (\ref{vorticity}) are integrated with double-precision using the fourth-order Runge-Kutta-Fehlberg adaptive scheme. The local error, relative to $\omega_{\max}(t)$, was kept below $10^{-10}$. The number of nodes $n = 1,\ldots,N$ was increased dynamically during the simulation in order to avoid the error due to truncation at small scales: the truncation error was kept below $10^{-20}$ for the enstrophy $\Omega = \frac{1}{2}\langle \omega_j , \omega_j \rangle$; see SM~\cite{SupMat} for more details. Together, this provided the remarkably high accuracy of numerical results. We stopped the simulation with $N = 80$, thus, covering the scale range of $\lambda^N \sim 10^{17}$ with the total of $13180$ time steps. The energy was conserved at all times with the relative error below $3.8\times 10^{-10}$.

\begin{figure*}[t]
\includegraphics[width=\textwidth]{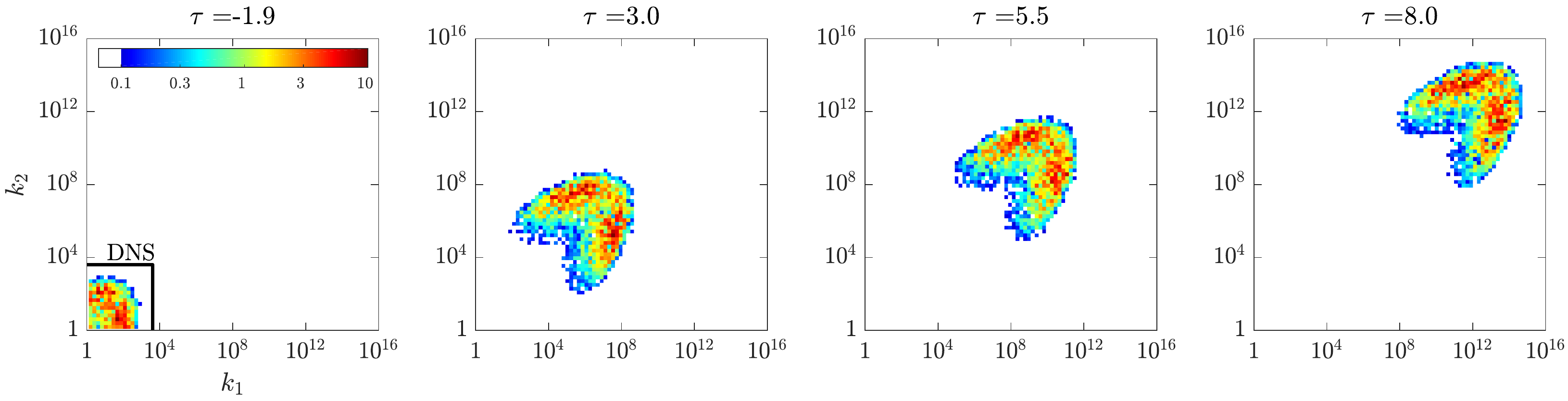}
\caption{Absolute value of the third component of renormalized vorticity $|\widetilde{\omega}_3|$, as a function of two positive wave numbers $k_1>0$ and $k_2>0$ (in log scale) at four different instants $\tau$. The third wave number is fixed at the node nearest to $k_3 = e^{\gamma\tau + 6} \propto (t_b-t)^{-\gamma}$. Values below $0.1$ are plotted in white. In the left figure, the small box bounds the square region $|k_{1,2}| \le 4096$, which would be accessible for the high-accuracy DNS with resolution $8192^3$. See also the Supplemental 3D video~\cite{SupMat}.}
\label{fig2}
\end{figure*}

Figures~\ref{fig1}(a) and~\ref{fig1}(b) analyze the temporal evolution of the maximum vorticity $\omega_{\max}(t) = \max_{\mathbf{k}\in\mathbb{\Lambda}^3}|\pmb{\omega}(\mathbf{k},t)|$ and the corresponding wave number $k_{\max}(t) = |\mathbf{k}|$. 
The Beale-Kato-Majda theorem~\cite{beale1984remarks} (whose proof for our model is identical to the continuous case) states that the blowup of the solution at finite time $t_b$ requires that the integral $\int_0^t \omega_{\max}(t)dt$ diverges as $t \to t_b$. In particular, this implies that the growth of maximum vorticity must be at least as fast as $\omega_{\max}(t) \gtrsim (t_b-t)^{-1}$. This dependence is readily confirmed in Fig.~\ref{fig1}(a) providing the blowup time $t_b = 15.870 \pm 0.001$. Furthermore, Fig.~\ref{fig1}(b) tracks the dependence $\omega_{\max}(t) \sim (t_b-t)^{-1}$ in logarithmic coordinates up to the values $\omega_{\max} \sim 10^{5}$. The same figure demonstrates the power-law dependence $k_{\max}(t) \sim (t_b-t)^{-\gamma}$ with the exponent $\gamma = 2.70\pm0.01$, simulated up to extremely small physical scales, $\ell \sim 1/k_{\max} \sim 10^{-15}$. Finally, Fig.~\ref{fig1}(c) shows the development of the power law $E(k) \propto k^{-\xi}$ in the energy spectrum as $t \to t_b$. The exponent can be obtained with the dimensional argument $E \propto \omega_{\max}^2/k_{\max}^3$, which yields $\xi = 3-2/\gamma \approx 2.26$.

\textsc{Chaotic blowup.}
The observed scaling agrees with the Leray-type~\cite{leray1934mouvement} self-similar blowup solution $\pmb{\omega}_L(\mathbf{k},t)$ defined as 
\begin{equation}
\pmb{\omega}_L(\mathbf{k},t) = (t_b-t)^{-1}\mathbf{W}[(t_b-t)^{\gamma}\mathbf{k}].
\label{Leray}
\end{equation}
Such a solution, however, cannot describe the blowup in Fig.~\ref{fig1}, where the maximum vorticity and the corresponding scale $\ell \sim 1/k_{\max}$ have the power-law behavior only in average, with persistent irregular oscillations. 

In order to understand the nonstationary blowup dynamics, we perform the change of coordinates 
\begin{align}
\widetilde{\pmb{\omega}} &= (t_b-t)\pmb{\omega}, & \eta &= \log|\mathbf{k}|, \nonumber \\
\mathbf{o} &= \mathbf{k}/|\mathbf{k}|, & \tau &= -\log(t_b-t).
\label{Renorm}
\end{align}
This change of coordinates applies similarly  in Fourier space $\mathbb{R}^3$ and in our 3D lattice $\mathbb{\Lambda}^3$. The Euler equations (\ref{vorticity}) in renormalized coordinates take the form
\begin{equation}
\partial_\tau \widetilde{\pmb{\omega}} = 
G[\widetilde{\pmb{\omega}}],
\label{eqRen}
\end{equation}
where the $i$th component of the nonlinear operator $G[\widetilde{\pmb{\omega}}]$ is 
\begin{equation}
\label{RHSrenorm}
(G[\widetilde{\pmb{\omega}}])_i = -\widetilde{\omega}_i -\widetilde{u}_j \ast \widetilde{\partial}_j\widetilde{\omega}_i + \widetilde{\omega}_j \ast \widetilde{\partial}_j \widetilde{u}_i, \quad \widetilde{\partial}_j = io_j;
\end{equation}
see SM~\cite{SupMat} for derivations. The choice of variables (\ref{Renorm}) is motivated by the scaling invariance: the operator $G[\widetilde{\pmb{\omega}}]$ is homogeneous (invariant to translations) with respect to $\tau$ and $\eta$, which correspond to temporal and spatial scaling, respectively. In our model, the scaling invariance is represented by the shifts of $\eta$ with integer multiples of $\log \lambda$. These properties allow studying the blowup as an attractor of system (\ref{eqRen}); see, e.g.,~\cite{eggers2009role,mailybaev2012renormalization}. For example, the self-similar blowup solution (\ref{Leray}) corresponds to the traveling wave $\widetilde{\pmb{\omega}} = \mathbf{W}(e^{\eta-\gamma\tau}\mathbf{o})$, which has a stationary profile in the comoving reference frame $\eta' = \eta-\gamma\tau$. In the limit $\eta \sim \gamma \tau \to \infty$, the original variables (\ref{Renorm}) yield the blowup dynamics: $|\pmb{\omega}| \to \infty$ and $\ell \sim 1/|\mathbf{k}| \to 0$ as $t \to t_b$. 
Such a blowup is robust to small perturbations if the traveling wave is an attractor in system (\ref{eqRen}).

\begin{figure}[b]
\includegraphics[width=0.81\columnwidth]{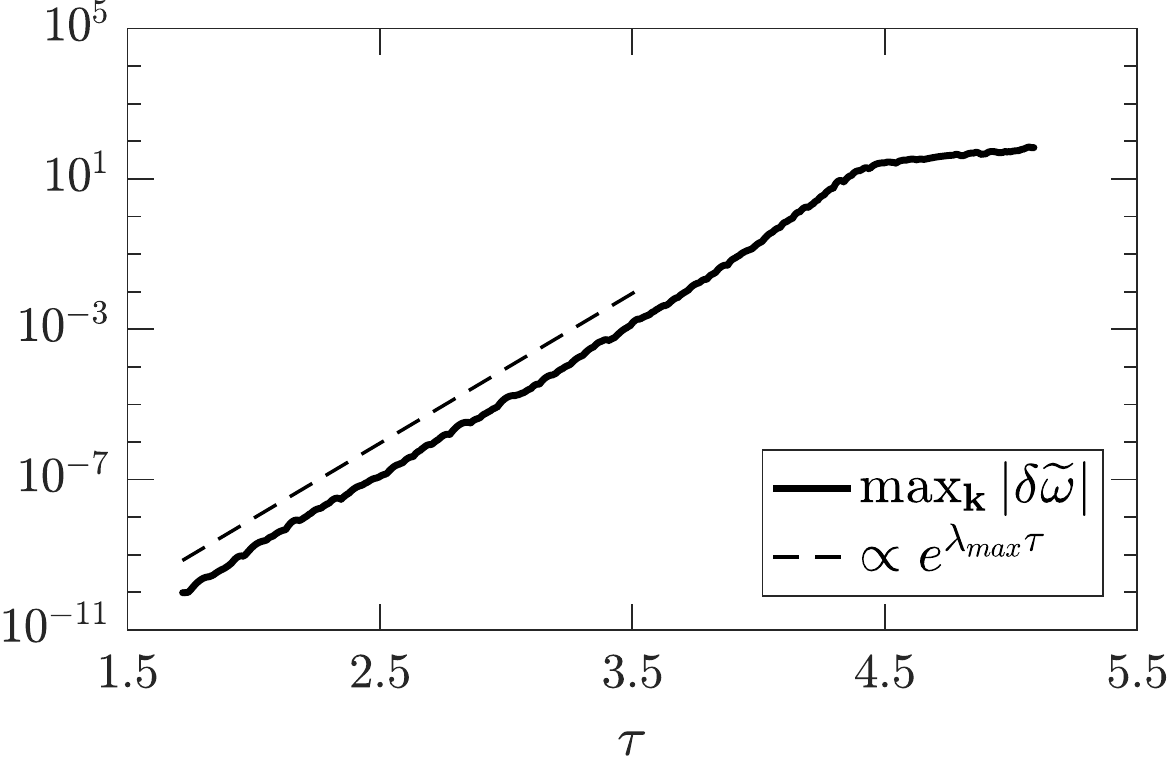}
\caption{Evolution of a small perturbation of vorticity, $\max_{\mathbf{k}}|\delta \widetilde{\pmb{\omega}}|$, in renormalized variables. Solutions deviate exponentially with the Lyapunov exponent $\lambda_{\max} \approx 9.18$.}
\label{fig3}
\end{figure}

Irregular evolution observed in Fig.~\ref{fig1} suggests that the attractor of system (\ref{eqRen}) cannot be a traveling wave. We will now argue that the attractor in the renormalized system represents a chaotic wave moving with the average speed $\gamma$. 
Figure~\ref{fig2} (see also the Supplemental video~\cite{SupMat} for the 3D picture) shows absolute values of the third component $\widetilde{\omega}_3$ as functions of two wave numbers $k_1 > 0$ and $k_2 > 0$ for four different values of $\tau$; here the third wave vector component is constant and chosen at the node nearest to $k_3 = e^{\gamma\tau + 6} \propto (t_b-t)^{-\gamma}$. This figure presented in log scale
demonstrates a wave moving with constant speed in average $\eta \sim \gamma \tau$, but not preserving exactly the spatial vorticity distribution. In order to confirm that the wave is chaotic, we computed the largest Lyapunov exponent $\lambda_{\max} = 9.18 \pm 0.07$ in Fig.~\ref{fig3}; here we added a tiny perturbation to the original solution at $\tau = 1.7$, when the attractor is already fully established, and observed the exponential deviation 
of the solutions $\max_{\mathbf{k}}|\delta \widetilde{\pmb{\omega}}(\tau)| \propto e^{\lambda_{\max}\tau}$ in renormalized time $\tau$. In the original variables, this yields the rapid power-law growth 
\begin{equation}
\max_{\mathbf{k}}|\delta \pmb{\omega}(t)| \propto (t_b-t)^{-\zeta},\quad
\zeta = \lambda_{\max}+1 \approx 10.18.
\label{lyapunov}
\end{equation}

\begin{figure}[t]
\includegraphics[width=\columnwidth]{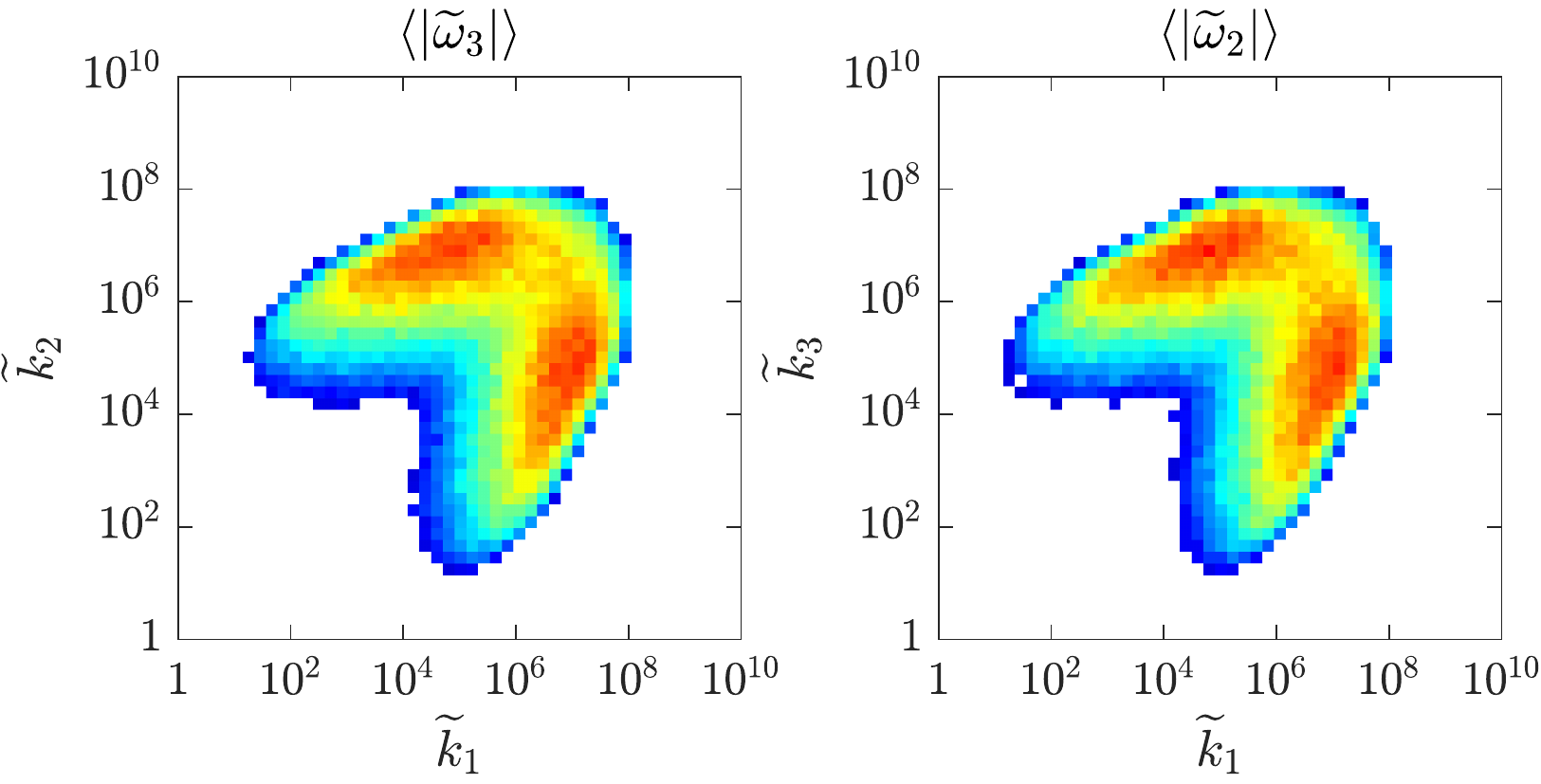}
\caption{Statistical isotropy: Left panel shows the $\tau$ average of $|\widetilde{\omega}_3|$ from Fig.~\ref{fig2} in a comoving reference frame $\eta' = \eta-\gamma\tau$. Right panel shows analogous result for the average of $|\widetilde{\omega}_2|$ on plane $(\widetilde{k}_1,\widetilde{k}_3)$. Planes of the two figures are related by the $90^\circ$ rotation about the $\widetilde{k}_3$ axis. Similar results are obtained for other elements of the rotation symmetry group $\mathsf{O_h}$.}
\label{fig4}
\end{figure}

The striking property of the chaotic attractor is that it restores the isotropy in the statistical sense, even though the solution at each particular moment is essentially anisotropic, in similarity to the recovery of isotropy in the Navier-Stokes turbulence~\cite{frisch1999turbulence,biferale2005anisotropy}. This property is confirmed in Fig.~\ref{fig4} presenting the averages of  renormalized vorticity components $|\tilde\omega_i|$, considered in the comoving reference frame $\eta'= \eta-\gamma\tau$. 
The isotropy, as well as other statistical properties, are expected to be established very rapidly in realistic conditions, e.g., in the presence of microscopic fluctuations, because of the very large Lyapunov exponent; see Eq.~(\ref{lyapunov}). This resembles closely a similar effect in developed turbulence~\cite{ruelle1979microscopic}.

\begin{figure}[t]
\includegraphics[width=0.88\columnwidth]{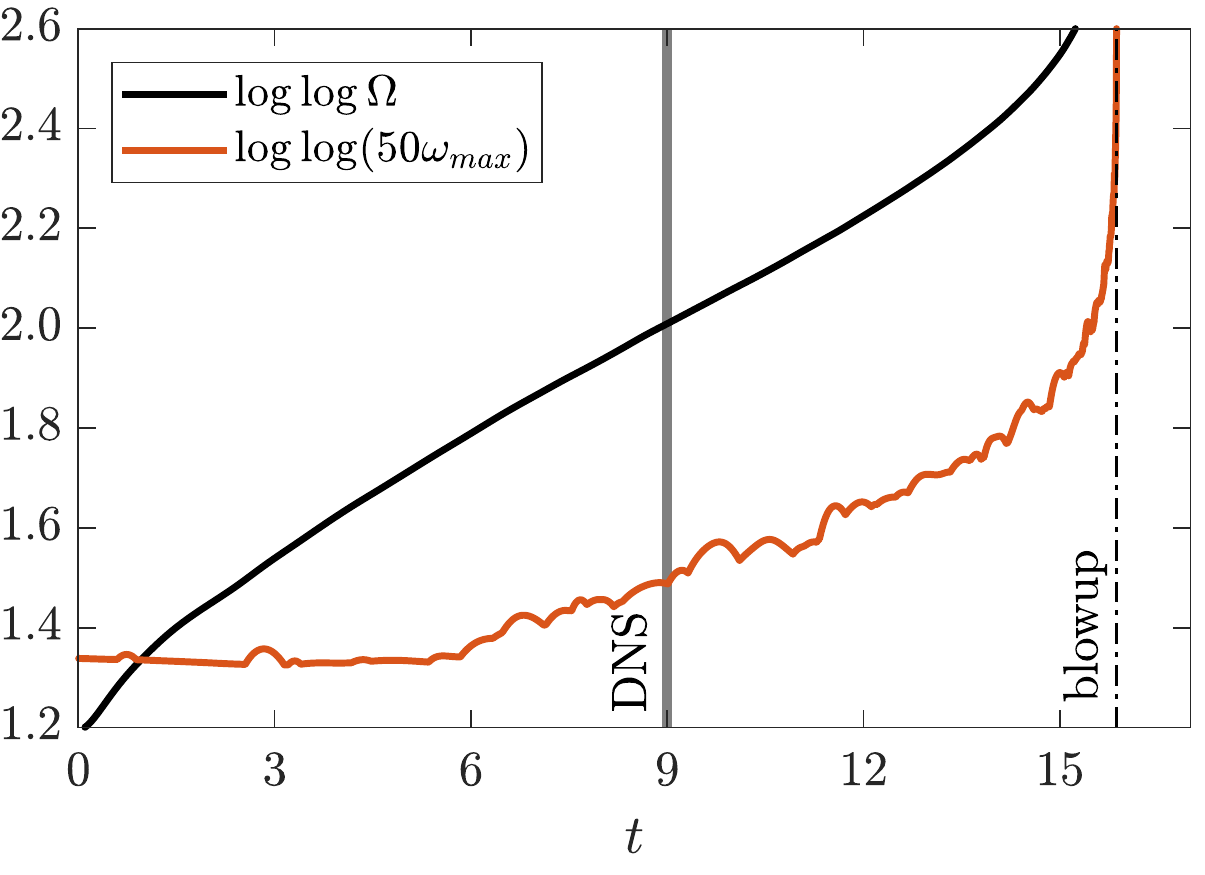}
\caption{Evolution of $\log\log\Omega$ and $\log\log\,(50\omega_{\max})$ for the enstrophy  and maximum vorticity; the factor $50$ is used to avoid complex values of the logarithm. The dash-dotted line indicates the blowup time. The vertical solid line estimates the limit $t_{\textrm{DNS}} \approx 9$ that would be accessible for the state-of-the-art DNS with the grid $8192^3$.  Until $t_{\textrm{DNS}}$, both $\Omega$ and $\omega_{\max}$ demonstrate the growth not greater than double exponential.}
\label{fig5}
\end{figure}

\textsc{Relation to existing DNS.} 
As one can infer from Figs.~\ref{fig2} and \ref{fig4}, the chaotic attractor has the span of about six decades of spatial scales. This property imposes fundamental limitations on the numerical resources necessary for the observation of blowup, assuming that the dynamics in the continuous 3D Euler equations can be qualitatively similar to our model. The approximate time limit, which would be accessible for the state-of-the-art DNS with the $8192^3$ grid \cite{grafke2008numerical,hou2009blow,kerr2013bounds} can be estimated in our model as $t_{\textrm{DNS}} \approx 9$ or $\tau_{\textrm{DNS}} \approx -1.9$ for the renormalized time; see Fig.~\ref{fig2} (left panel). At this instant, the chaotic attractor is still at its infant formation stage and, hence, the dynamics is essentially transient. 
The increase of the vorticity from $\omega_{\max}(0) = 0.91$ to $\omega_{\max}(t_{\textrm{DNS}}) = 1.89$ and of the enstrophy from $\Omega(0) = 27.2$ to $\Omega(t_{\textrm{DNS}}) = 1.92 \times 10^3$ is moderate, which is also common for the DNS. 
Moreover, Fig.~\ref{fig5} shows that the growth of enstrophy and vorticity for $t \lesssim t_{\textrm{DNS}}$ is not faster than double exponential in agreement with~\cite{hou2007computing,hou2009blow,kerr2013bounds}. The chaotic blowup behavior offers a  diversity of flow structures as it is indeed observed for different initial conditions~\cite{gibbon2008three}; some DNS showed the incipient development of power-law energy spectra~\cite{agafontsev2015}, in qualitative agreement with Fig.~\ref{fig1}(c).

At the time $t_{\textrm{DNS}}$, the wave vector at the vorticity maximum is equal to $\mathbf{k}_{\max} = (\lambda^6,-\lambda^3,\lambda^{10}) \approx (18.9,-4.2,123)$. Its third component is much larger than the other two. This has a similarity with DNS, 
which typically demonstrate depleting of vorticity growth within quasi-2D (thin in one and extended in the other two directions) vorticity structures~\cite{brachet1992numerical,frisch2003singularities,agafontsev2017asymptotic}. Such dominance of one scale over the others by 1 or 2 orders of magnitude persists for larger times in our model. 

\textsc{Conclusions.}
We propose an explanation for the existing controversy in the blowup problem for incompressible 3D Euler equations. This is accomplished using a new model, which is formally identical to the incompressible Euler equations and defined on the 3D logarithmic grid with proper algebraic operations. Such a model retains most symmetries of the original system along with intrinsic invariants (energy, helicity, circulation, etc.), but permits simulations in extremely large interval of scales. 

We show that our model has the non-self-similar blowup, which is explained as a chaotic attractor in renormalized equations. 
Our results demonstrate that the blowup has enormously higher complexity than anticipated before: its ``core'' extends to six decades of spatial scales. This suggests that modern DNS of the original continuous model are unsuitable, by far, for the blowup observation; still, the blowup may be accessible to experimental measurements. 
Since the attractor is chaotic, blowup cannot be probed by the study of local structures. 

Our approach to the blowup phenomenon is not limited to the Euler equations, but is ready-to-use for analogous studies in other fields such as natural convection, geostrophic motion, magnetohydrodynamics, and plasma physics.

The authors are grateful to Luca Biferale, Gregory Eyink, Uriel Frisch, Simon Thalabard and Dmitry Agafontsev for most helpful discussions. The work was supported by the CNPq Grant No. 302351/2015-9 and the RFBR Grant No. 17-01-00622.

\bibliography{refs}

\section{Supplemental material}

\subsection{Conservation laws}

Conservation of quadratic invariants follows from the fact that the product \eqref{convolution} contains only the exact triples of wave vectors. Taking the energy as an example, let us show how the proof can be written using the basic operations defined on the 3D logarithmic lattice, following the standard approach of fluid dynamics.
Using the Euler equations \eqref{euler}, we obtain
\begin{equation}
\begin{array}{rcl}
\displaystyle
\frac{dE}{dt} &=& 
\displaystyle
\frac{d}{dt} \left( \frac{1}{2} \langle u_i,u_i \rangle \right) = \langle u_i,\partial_t u_i \rangle \\[12pt] 
&=& -\,\langle u_i,\partial_i p \rangle -\langle u_i,u_j \ast \partial_j u_i \rangle.
\end{array}
\end{equation}
The pressure term vanishes owing to the incompressibility condition as
\begin{equation}
\label{byparts}
\langle u_i,\partial_i p \rangle = -\langle \partial_i u_i,p \rangle = 0,
\end{equation}
where the first relation represents the derivation by parts on the 3D lattice.
In the inertial term, using commutativity of the product and the properties 
\eqref{leibniz} and \eqref{associativity}, one obtains
\begin{align*}
\langle u_i,u_j \ast \partial_j u_i \rangle 
= \langle u_i \ast \partial_j u_i, u_j \rangle = \frac{1}{2} \langle \partial_j (u_i \ast u_i),u_j \rangle.
\end{align*}
After integration by parts, analogous to \eqref{byparts}, this term vanishes due to  the  incompressibility condition.

Conservation of helicity can be proved following a similar line of derivations. For the Beale-Kato-Majda theorem, one has to define the functional spaces and the corresponding inequalities; technical details of this functional analysis on the 3D logarithmic lattice will be given elsewhere.

Furthermore, one can make sense of Kelvin's circulation theorem in system \eqref{euler}. It is related to the conservation of cross-correlation $\Gamma = \langle u_j,h_j \rangle$ for an arbitrary ``frozen-into-fluid'' divergence-free field $\mathbf{h}(\mathbf{k},t) = (h_1,h_2,h_3)$ satisfying the equations~\cite{moffatt1969degree} 
\begin{equation}
\partial_t h_i + u_j \ast \partial_j h_i - h_j \ast \partial_j u_i = 0, \quad \partial_j h_j = 0.
\label{frozen_field}
\end{equation}
In the continuous formulation, the circulation around a closed material contour $\mathbf{C}(s,t)$ in physical space ($s$ is the arc length parameter) is given by the  cross-correlation $\Gamma$ with the field $\mathbf{h}(\mathbf{x},t) = \oint \frac{\partial\mathbf{X}}{\partial s}\,\delta^3(\mathbf{x}-\mathbf{C}(s,t))\,ds$, where $\delta^3$ is the 3D Dirac delta-function; see, e.g. \cite{zakharov1997hamiltonian,majda2002vorticity}. Therefore, $\Gamma$ represents the generalized circulation in Kelvin's theorem. Its conservation yields the infinite number of circulation invariants in our model: the cross-correlation $\Gamma$ is conserved for any solution of system \eqref{frozen_field}.

Note that zero wave number can also be considered in the model by adding it into the set $\mathbb{\Lambda}$. This will further increase the number of terms in the sum \eqref{convolution}.
Note also that the number of degrees of freedom in our model is substantially smaller than in the original Euler system. On one hand, this is an important advantage of the model that allows numerical simulations for extremely large range of scales. On the other hand, one should be careful when using this model in situations where thermalization at small scales may play a role, e.g.,~\cite{cichowlas2005effective,bowman2006links}.

\subsection{Initial conditions}

Initial conditions used in numerical simulations are given below in terms of velocities. Nonzero components are limited to large scales $\lambda \le |k_{1,2,3}| \le \lambda^3$ and taken in the form
\begin{equation}
u_j(\mathbf{k}) = \frac{|\epsilon_{jmn}|}{2}k_m k_n e^{i\theta_j(\mathbf{k})-|\mathbf{k}|}, \quad \text{for} \quad j=1,2.
\end{equation}
Here $\epsilon_{jmn}$ is the Levi-Civita permutation symbol and the phases $\theta_j$ are given by
\begin{equation}
\begin{array}{rcl}
\theta_j(\mathbf{k}) & = & \mathrm{sgn}(k_1)\alpha_j + \mathrm{sgn}(k_2)\beta_j
+\mathrm{sgn}(k_3)\delta_j
\\[3pt]
&&  +\,\mathrm{sgn}(k_1k_2k_3)\gamma_j 
\end{array}
\end{equation}
with the constants $(\alpha_1,\beta_1,\delta_1,\gamma_1) = (1,-7,13,-3)/4$ and $(\alpha_2,\beta_2,\delta_2,\gamma_2) = (-1,-3,11,7)/4$. The third component of velocity is uniquely defined by the incompressibility condition. Several tests were also performed with random initial conditions limited to large scales. In all the test, we observed the same chaotic attractor of the renormalized system and, therefore, the same (universal) asymptotic form of the chaotic blowup.

\subsection{Adaptive scheme}

Since only a finite number of modes can be simulated, the infinite-dimensional nature of the problem was tracked very accurately by using the following adaptive scheme in the simulation. At each time step, we computed the enstrophy of the modes with the wave numbers $|\mathbf{k}| \ge K_{\max}/\lambda$, where $K_{\max}$ is the largest wavenumber in each direction of the lattice. This quantity estimates the enstrophy error due to mode truncation, and it was kept extremely small, below $10^{-20}$, during the whole simulation. Every time the threshold of $10^{-20}$ was reached we increased the number of nodes in each direction by five, i.e., multiplying $K_{\max}$ by $\lambda^5$.

\subsection{Renormalized Euler equations}

With the renormalized variables \eqref{Renorm}, it is convenient to define new differentiation operators as the Fourier factors $\widetilde{\partial}_j = io_j$, where $\mathbf{o} = (o_1,o_2,o_3) = \mathbf{k}/|\mathbf{k}|$ and $i$ is the imaginary unit. Thus, derivatives in the original and in the renormalized variables are related as $\partial_j = e^{\eta}\widetilde{\partial}_j$. Also, the renormalized velocity can be defined as
$\widetilde{\mathbf{u}} = (t_b-t)|\mathbf{k}|\mathbf{u}$,
which is related to the renormalized vorticity as
\begin{equation}
\widetilde{\mathbf{u}} = i\mathbf{o} \times \widetilde{\pmb{\omega}}. 
\label{renorm_u}
\end{equation}

Using relations \eqref{Renorm} and \eqref{renorm_u}, the vorticity equation \eqref{vorticity}, after dropping the common factor $e^{2\tau}$, becomes
\begin{equation}
\partial_{\tau}\widetilde{\omega}_i + \widetilde{\omega}_i 
+ \widetilde{u}_j \ast \widetilde{\partial}_j\widetilde{\omega}_i - \widetilde{\omega}_j \ast \widetilde{\partial}_j \widetilde{u}_i = 0.
\end{equation}
This equation has the form \eqref{eqRen}--\eqref{RHSrenorm}.
Since $\tau$ and $\eta$ do not appear explicitly in \eqref{RHSrenorm}, the renormalized system (8) is translation invariant with respect to these two variables.

Note that the existence of a chaotic wave traveling with constant mean velocity in the renormalized system yields the power law $\omega_{\max}(t) \sim (t_b-t)^{-1}$ observed in Fig.~\ref{fig1}(b). This follows from the transformation \eqref{Renorm}, similarly to the Leray-type solution \eqref{Leray}. In fact, existence of a chaotic or regular wave with a positive speed $\gamma$ as an attractor in the renormalized system is a sufficient condition for the finite-time blowup.

\end{document}